\documentclass[conference,a4paper]{APSIPA2025}
\usepackage{amsmath}
\usepackage{graphicx}
\usepackage{multirow}
\usepackage{threeparttable}
\usepackage[backend=biber,style=ieee,]{biblatex}
\usepackage{booktabs}   
\usepackage{tabularx}   
\usepackage{graphicx}   
\usepackage{siunitx} 
\usepackage{textcomp}

\usepackage{geometry}
\usepackage{fancyhdr}

\fancypagestyle{firststyle}{
  \fancyhf{}
  \fancyhead[C]{2025 Asia Pacific Signal and Information Processing Association Annual Summit and Conference (APSIPA ASC)}
}

\begin{document}

\title{Improving Code-Switching Speech Recognition with TTS Data Augmentation}

\author{
\authorblockN{
Yue Heng Yeo\authorrefmark{1}\authorrefmark{2}, 
Yuchen Hu\authorrefmark{2},
Shreyas Gopal\authorrefmark{2},
Yizhou Peng\authorrefmark{2}, 
Hexin Liu\authorrefmark{2},
and Eng Siong Chng\authorrefmark{2}
}

\authorblockA{
\authorrefmark{1}
Institute for Infocomm Research (I$^2$R), A*STAR, Singapore}

\authorblockA{
\authorrefmark{2}
College of Computing and Data Science, Nanyang Technological University, Singapore \\
E-mail: yueheng001@ntu.edu.sg}
}
\pagestyle{plain}
\maketitle
\thispagestyle{firststyle}

\begin{abstract}
Automatic speech recognition (ASR) for conversational code-switching speech remains challenging due to the scarcity of realistic, high-quality labeled speech data. This paper explores multilingual text-to-speech (TTS) models as an effective data augmentation technique to address this shortage. Specifically, we fine-tune the multilingual CosyVoice2 TTS model on the SEAME dataset to generate synthetic conversational Chinese–English code-switching speech, significantly increasing the quantity and speaker diversity of available training data. Our experiments demonstrate that augmenting real speech with synthetic speech reduces the mixed error rate (MER) from 12.1\% to 10.1\% on DevMan and from 17.8\% to 16.0\% on DevSGE, indicating  performance gains. These results confirm that multilingual TTS is an effective and practical tool for enhancing ASR robustness in low-resource, conversational code-switching scenarios.
\end{abstract}

\section{Introduction}
Code-switching is an everyday practice where multilingual speakers mix two or more languages into a single conversation, in either intra-sentence or inter-sentence manner, choosing words or grammatical structures that best fit their communicative intent~\cite{auer1999codeswitching, bullock2009cambridge, liu21xsa}. In automatic speech recognition, code-switching is particularly challenging because speakers often adjust their intonation, rhythm, and pronunciation when transitioning between languages, demanding ASR systems to track these shifts in real time~\cite{li2013spoken, vu2012first, liu24align, liu2023enhancing}. Despite the existing advancements~\cite{li2021data, lyu2015seame, winata2022codeswitch,liu23_icassp, mamba_in_speech}, a significant obstacle for code-switching ASR is the scarcity of realistic, accurately transcribed code-switching datasets, severely limiting model performance.

A prevalent method to address the shortage of code-switching data is audio splicing~\cite{hussein2023speechcollage}. This technique synthesizes CS speech by concatenating audio segments from separate monolingual recordings, creating synthetic bilingual utterances without additional data collection. Empirically, ASR systems trained on audio-spliced data have demonstrated improvements in error rates and reduced monolingual bias. However, concatenating audio segments typically results in unnatural prosody and noticeable acoustic discontinuities, introducing co-articulation artifacts that may lead to model overfitting. Consequently, despite being useful for initial experimentation, audio-spliced data exhibits inherent limitations in realism and linguistic coverage compared to advanced TTS-generated synthetic speech, particularly in conversational datasets such as SEAME~\cite{winata2022codeswitch, nguyen2025codeswitch}.

Another possible method is TTS augmentation. While early TTS models were considered ineffective for code-switching ASR due to difficulties in modeling natural prosody, speaker variability, and complex language-switching patterns~\cite{rosenberg2019speech, winata2022codeswitch, zhang2024speaking}, recent work by Chou et al.~\cite{chou2025selfrefining} demonstrates that synthetic speech generated by advanced TTS models can significantly improve ASR performance. Their self-refining framework, which leverages TTS-synthesized data, achieves significant reduction in error rates ASR task, highlighting the practical effectiveness of TTS augmentation for fine-tuning ASR systems in code-switching scenarios.

Identified key factors for successful augmentation include sufficient text diversity, moderate speaker variation, and appropriate balance between real and synthetic speech \cite{yang2024enhancing, li2021data}. Leveraging such versatile TTS models to generate synthetic data provides an effective and cost-efficient solution because it bypasses the expensive stages of speaker recruitment, studio recording, and manual code-switch transcription by relying solely on readily crawled text and automatically self-labelled speech embeddings to augment real-world code-switching datasets, directly addressing data scarcity \cite{yang2024enhancing, winata2022codeswitch}. Recent studies have demonstrated substantial performance improvements when ASR systems are trained using synthetic speech generated by advanced multilingual TTS models, significantly reducing the performance gap relative to real-world data \cite{yang2024enhancing, zhang2022synthetic}.

\begin{figure*}[t]
  \centering
  \includegraphics[width=0.8\textwidth]{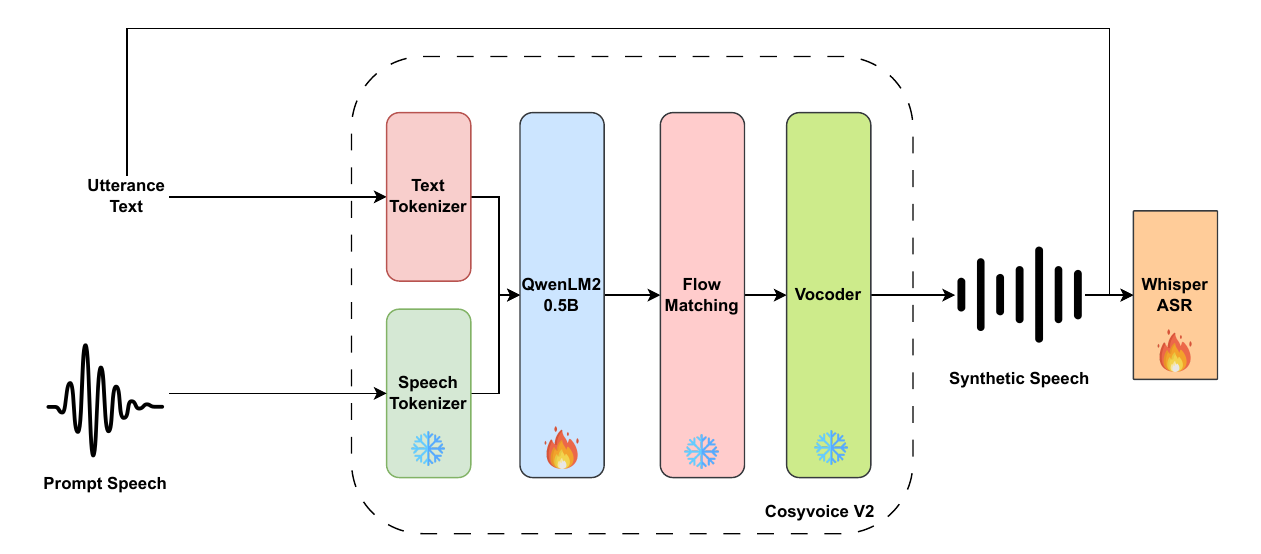}
  \caption{End-to-end synthetic-data pipeline.  Ground-truth text and speech are tokenised, passed through the Qwen-2 language model, a
   flow-matching decoder and a HiFT vocoder to yield synthetic audio,
   which is later used to fine-tune Whisper.}
  \label{fig:cosyvoice_pipeline}
\end{figure*}

\section{Related Works}
Recent advances in ASR increasingly utilize synthetic speech generated by TTS systems to alleviate the data scarcity challenges in low-resource multilingual scenarios~\cite{park2019specaugment, rosenberg2019speech, li2021data}. Yang \textit{et al.}~\cite{yang2024enhancing}, for instance, demonstrated significant ASR performance gains by leveraging the multilingual CosyVoice-base TTS model across diverse low-resource domains such as accented speech, minority languages (Korean, Chinese dialects), and specialized vocabulary (e.g., automotive hotwords). Their results emphasized the importance of adequate textual and moderate speaker diversity for effective TTS augmentation.

Nevertheless, while synthetic TTS speech augmentation has proven effective in various linguistic contexts, most prior research has overlooked conversational code-switching, which introduces unique challenges like rapid intra-sentence language switching, informal lexical usage, and complex prosodic patterns~\cite{winata2022codeswitch, bullock2009cambridge}. Common prior approaches, such as audio-splicing augmentation by concatenating segments of monolingual audio to form synthetic bilingual utterances, have achieved modest improvements but often produce unnatural prosody and acoustic artifacts, limiting their effectiveness in realistic conversational scenarios~\cite{li2021data, zhang2022synthetic}.

To address these limitations, our study explicitly investigates using the modern multilingual CosyVoice TTS model, fine-tuned specifically on conversational Chinese--English code-switching corpora such as SEAME~\cite{lyu2015seame}. By emphasizing realistic conversational structure, spontaneous prosodic variation, and diverse speaker characteristics, we aim to substantially enhance ASR system robustness within complex code-switching conversational contexts, thereby extending existing multilingual augmentation frameworks.

\subsection{Our Contributions}
Our contributions in this paper are summarised as follows:
\begin{itemize}
\item We demonstrate that multilingual TTS models, specifically CosyVoice fine-tuned on code-switching datasets, effectively capture realistic conversational prosody, informal lexical usage, and rapid intra-sentence language switching. Furthermore, showing that the TTS data is possible for finetuning speech foundation models.

\item We identify critical factors for successful TTS augmentation, by adding speaker variation, adding amount of data, and an optimal balance between synthetic and real speech data.

\item We confirm the adaptability and effectiveness of our TTS augmentation pipeline by successfully transferring a SEAME-fine-tuned CosyVoice model to a different code-switching corpus (ASCEND), greatly narrowing the gap with real data performance.
\end{itemize}

\section{Methods}
\subsection{CosyVoice TTS}
CosyVoice~\cite{du2024cosyvoice} is a scalable, multilingual, zero-shot text-to-speech (TTS) system that relies on supervised semantic tokens extracted from a multilingual ASR encoder. As summarised in Figure 1, the architecture is organised into four tightly coupled blocks. (1) Text encoder: a language-agnostic BPE front-end converts the input sentence into tokens and aligns them to the speech timeline. (2) Speech tokenizer: built on vector-quantisation over the ASR encoder, this module discretises training audio into low-rate semantic codes. (3) Large language model (LLM) the backbone of the TTS system: treating TTS as an autoregressive sequence generation task, a transformer-based LLM takes in the mixed stream of text and speech tokens and autoregressively predicts the next speech token based on the inputted tokens. (4) Conditional flow-matching decoder: the generated token sequence is up-sampled and passed through a flow-matching network that converts it into mel-spectrograms, which a lightweight vocoder renders as waveform.

CosyVoice is trained in two stages: the speech tokenizer learns from  approximately 200k hours of aligned Chinese-English audio, while the full TTS model is trained on an additional 167k hours spanning four languages: Chinese, English, Japanese and Korean. This scale and diversity allow CosyVoice to generate natural, speaker-consistent speech, including fluent intra-sentence code-switching, making it a practical source of synthetic data for strengthening multilingual ASR systems.
\subsection{Whisper ASR}
Whisper is a large-scale Transformer-based ASR (ASR) model from OpenAI\footnote{Model at \url{https://github.com/openai/whisper}.}, trained on about 680k hours of multilingual audio. Because its training data spans a broad range of acoustic conditions and spoken languages, Whisper often excels in handling diverse speakers, accents, and noisy recordings.

However, code-switching remains problematic. Although Whisper’s multilingual approach can usually handle multiple languages independently, it can struggle when they appear in rapid alternation, leading to transcription errors or incorrect language identification.

\subsection{Data Generation Pipeline}
Our method has three stages, to data augment reference speech and reproduce more data in terms of speaker variety and volume of data.

\subsubsection {TTS tuning}
We begin by adapting CosyVoice 2 to the SEAME domain.
Only the QwenLM language-model component is updated; the speech
tokeniser, flow-matching decoder and vocoder are kept fixed.
During fine-tuning, QwenLM is trained to auto-regressively predict
speech-token sequences given SEAME text tokens, allowing it to internalise the corpus’s rapid Mandarin–English alternations, informal phrasing and conversational prosody.
This single-module update is computationally lightweight yet
sufficient to steer the TTS system toward natural code-switched output
while preserving the acoustic fidelity of the original CosyVoice stack.

\subsubsection {Synthetic speech generation}
After adaptation, each SEAME transcript is re-synthesised multiple
times using different x-vector speaker embeddings sampled from a large
pool.
The result is a speaker-diverse synthetic corpus that mirrors the
original text but enriches timbre, pitch range and speaking-rate
variation.

\subsubsection {ASR tuning}
The synthetic speech is mixed with the 100 h SEAME ground-truth audio
and used to fine-tune Whisper-small model.
We compare three conditions: (i) Ground Truth-only, (ii) Ground Truth + TTS (the proposed mix) and (iii) TTS-only.
Keeping the Whisper architecture and augmentation recipe unchanged
lets us isolate the impact of the additional, speaker-rich synthetic
data on code-switching recognition accuracy.

\section{Experiment Setup}
\subsection{SEAME Dataset}
The SEAME (South-East Asia Mandarin-English) corpus is a speech dataset designed specifically to capture spontaneous, conversational code-switching between Mandarin and English among bilingual speakers in Singapore and Malaysia. It contains approximately 192 hours of audio from natural conversations and interviews involving 156 speakers. Conversations cover everyday topics and showcase frequent switches between languages, often even within a single sentence or phrase. Each utterance is carefully transcribed, with clear labels marking language boundaries, making SEAME ideal for training and evaluating automatic speech recognition systems that must handle real-world bilingual interactions. Due to its spontaneous nature, realistic language mixing, and detailed annotations, SEAME is widely used as a standard benchmark for code-switching research and development.

\subsection{Model}
\label{sec:exp_setup}

\paragraph{CosyVoice fine-tuning}
We adapt the CosyVoice 2's QWENLM2 (0.5 B parameters) model to the target domain.  Optimisation uses Adam with an initial learning rate of \(\mathit{1\times10^{-4}}\). The rate grows linearly during the first 10 000 updates (warm-up) and then
remains constant for the rest of the 200 training epochs.

\paragraph{Whisper ASR fine-tuning.}
The ASR back-end starts from the released Whisper-small checkpoint for the abalation studies(\(\sim\)240 M parameters) and is fine-tuned in ESPnet \footnote{\url{https://github.com/espnet/espnet}}.
Input waveforms are converted to 80-bin log-Mel filter-banks
(24 kHz, 20 ms window, 12 ms hop); we apply the same SpecAugment
two frequency masks (width $\le 40$ bins), five time masks
(width $\le 12\,\%$ of the utterance) and a five-frame time-warp window.
Optimisation uses AdamW (\(\beta = 0.9/0.99\), \(\epsilon = 1\times10^{-6}\), weight decay 0.01).
The learning rate follows ESPnet’s \texttt{warmuplr} schedule:
it rises linearly from zero to $1\times10^{-5}$ over the first
1 500 updates, then decays with the inverse-square-root rule.
Mini-batches are built by counting the total number of spectrogram elements;
each update is limited to about 12 M elements, with gradients accumulated
over four steps. The language id has been set auto for all experiments.

\paragraph{Evaluation.}
ASR quality is reported as mixed-error rate (MER) on \textsc{DevMan} and \textsc{DevSGE} inside ESPNET toolkit for SEAME recipe.

\subsection{Experiment Results}

\begin{table}[t]  
  \centering
  \caption{Mixed-Error Rate (MER) of Whisper-Largev3 on DevMan and DevSGE for different mixes
           of real speech, original-speaker TTS (TTS-O), and random-speaker
           TTS (TTS-R).  Bold marks the lowest MER in each test set.}
  \begin{tabularx}{\columnwidth}{c|ccc|cc}
    \toprule
    \textbf{Model} &
    \multicolumn{3}{c|}{\textbf{Duration (h)}} &
    \multicolumn{2}{c}{\textbf{MER (\%)}} \\
    \cmidrule(lr){2-4}\cmidrule(l){5-6}
    & \textbf{Real} & \textbf{TTS-O} & \textbf{TTS-R} &
      \textbf{DevMan} & \textbf{DevSGE} \\
    \midrule
    \multirow{6}{*}{Whisper-Largev3}
      & 100 & -   & -   & 12.1 & 17.8 \\
      & -   & 100 & -   & 12.5 & 18.6 \\
      & -   & -   & 100 & 17.7 & 22.4 \\
      & 100 & 100 & -   & 11.1 & 17.0 \\
      & 100 & -   & 100 & \textbf{10.1} & \textbf{16.0} \\
      & -   & -   & 200 & 12.2 & 18.5 \\
    \bottomrule
  \end{tabularx}
  \label{tab:real_vs_tts_split}
\end{table}

The results in Table \ref{tab:real_vs_tts_split} confirm that high-quality TTS is an effective data-augmentation tool for Whisper-Largev3 when synthetic speech is added in addition to the available real recordings. Our baseline of fine-tuning Whisper with only the 100 h of real speech yields 12.1\% MER on DevMan and 17.8\% on DevSGE. Regenerating those same utterances with CosyVoice2 while keeping the original speaker embeddings (TTS-O) and mixing the two sets one-to-one reduces MER to 11.1\% / 17.0\%. The most substantial benefit appears when replacing speaker embeddings with randomly sampled ones (TTS-R): combining 100 h of real speech with 100 h of random-speaker synthesis lowers MER further to 10.1\% on DevMan and 16.0\% on DevSGE, surpassing the ground-truth baseline by roughly two absolute points on each test set. The pattern suggests the crucial ingredient is speaker diversity; synthetic audio that merely repeats original voices contributes less than audio introducing new timbres and prosodies. At the same time, training on 200 h of random-speaker TTS alone underperforms the real-only model (12.2\% DevMan, 18.5\% DevSGE), indicating synthetic data works best as a complement rather than a substitute. Taken together, these findings show that TTS can deliver “almost real” training examples that substantially improve recognition accuracy, provided the synthetic set at least doubles the real-data volume and introduces fresh speaker characteristics rather than duplicating existing ones.
\subsection{Amount of Data to finetune Cosyvoice}
\begin{table}[t]
  \centering
    \caption{MER (\%) of Whisper-small on DevMan and DevSGE when trained on synthetic
           speech produced by a TTS model fine-tuned with different amounts
           of target-domain data.  UTMOS denotes the mean opinion score of the
           corresponding synthetic sets; bold marks the best value in each
           column.}
  \begin{tabular}{@{}cccc@{}}
    \toprule
    \textbf{Duration (h)} &
    \textbf{DevMan} &
    \textbf{DevSGE} &
    \textbf{UTMOS} \\
    \midrule
    Ground-Truth & \textbf{13.4} & \textbf{19.2} & \textbf{3.6} \\
    \midrule
    10       & 21.8          & 26.1          & 2.9 \\
    50       & 15.2          & 23.2          & 3.1 \\
    100      & 13.8          & 20.1          & 3.2 \\
    \bottomrule
  \end{tabular}

  \label{tab:cv_ft_pivot}
\end{table}

To find out how much data we need to finetune Cosyvoice2 to replicate more in-domain data, we ran UTMOS22~\cite{saeki2022utmos}, an open-source model that predicts a mean-opinion score (MOS) from 1 (poor) to 5 (excellent).
The real recordings (ground truth) reach 3.6. When CosyVoice is fine-tuned on reach 10 h of target speech, the MOS reaches to 2.9. Expanding the fine-tune pool to 50 h raises the score to 3.1, and using the full 100 h nudges it to 3.2. The MOS curve climbs only modestly because CosyVoice had already been pre-trained on hundreds of hours of multi-speaker data. The large-scale pre-training taught the model most of the clear pronunciation, smooth pitch, and low noise so even the 10 h version starts from a reasonably high baseline. Extra in-domain hours mainly help the TTS copy the conversational style and rhythm of our corpus, which big MOS models reward only slightly. In contrast, the ASR metrics respond much more: the same jump from 10 h to 100 h cuts MER by roughly nine absolute points. Thus, while MOS gains are little, the larger fine-tune sets remain valuable because they push the synthetic speech closer to the target domain of multi-turn conversation codeswitching speech in ways that matter for Whisper-small’s recognition accuracy.
\footnote{\url{https://github.com/sarulab-speech/UTMOS22}}
\subsection{Amount of Data to Synthesise to finetune Whisper}
\begin{table}[t]
  \centering
    \caption{MER (\%) on DevMan and DevSGE as Whisper-small is
           fine-tuned with increasing amounts of synthetic speech.
           The last column shows the relative MER reduction obtained by
           adding each extra 100-h block (averaged over both test sets).}
  \begin{tabular}{@{}cccc@{}}  
    \toprule
    \textbf{Synthetic Data (h)} &
    \textbf{DevMan} &
    \textbf{DevSGE} &
    \textbf{Rel. Gain (\%)} \\
    \midrule
    Ground-Truth & 13.4 & 19.2 & -- \\   
    \midrule
    100 & 19.0 & 23.7 & --    \\
    200 & 13.3 & 19.9 & \textbf{23.04} \\
    300 & 11.7 & 18.2 & 10.29 \\
    400 & 11.2 & 17.5 & 4.06 \\
    500 & \textbf{10.9} & \textbf{17.0} & 2.54 \\
    \bottomrule
  \end{tabular}
  \label{tab:MER_vs_hours}
\end{table}
Table \ref{tab:MER_vs_hours} shows that enlarging the synthetic set beyond the 100-hour ground-truth baseline consistently drives MER down, but the marginal benefit shrinks with each additional block of data. Doubling the training hours to 200 h delivers the most substantial payoff: mixed MER falls by roughly one-quarter compared with the baseline, making this step the single largest improvement in the series. A further rise to 300 h still helps, but the extra gain is only about half of what the previous increment provided. Beyond that point the curve flattens sharply. The 400 h and 500 h conditions shave off just four and three tenths of a percentage point, respectively while incurring the full cost of another one-hundred-hour synthesis run. These results indicate that the ideal amount of data to finetune for Whisper-small lies between two and three times the amount of real data: it captures most of the performance upside without sliding into the zone of rapidly diminishing returns.
\subsection{Data Augmentation Comparison}
\begin{table}[t]
  \centering
    \caption{MER (\%) of Whisper-small after 300 h of fine-tuning with two
           data-augmentation techniques (lower is better; best scores in bold).}
  \begin{tabular}{@{}lcc@{}}   
    \toprule
    \textbf{Technique} &
    \textbf{DevMan} &
    \textbf{DevSGE} \\
    \midrule
    GT + Speed perturbation & 13.4 & 19.2 \\
    GT + TTS                & \textbf{12.6} & \textbf{18.7} \\
    \bottomrule
  \end{tabular}

  \label{tab:MER_technique}
\end{table}
Another common speech augmentation technique to train ASR systems is speed perturbation. Speed perturbation is a widely used audio‐augmentation method that synthetically varies speech tempo by resampling each waveform at a small set of fixed scaling factors typically 0.9 × (slower), 1.0 × (original), and 1.1 × (faster). Because the operation stretches or compresses the time axis while leaving the spectral envelope largely intact, it preserves the speaker’s timbre and linguistic content yet introduces realistic rate-of-speech variability. Applying all three factors effectively triples the amount of training data without requiring additional transcription, providing the acoustic model with broader coverage of temporal dynamics and improving robustness to speaking-rate mismatches at test time.\par
Table~\ref{tab:MER_technique} contrasts the two similar data augmentation technique fine-tuning for \emph{Whisper-small}. 
Comparing between the conventional three-speed perturbation (\mbox{0.9/1.0/1.1}$\times$) with a 300-hour CosyVoice TTS augmentation lowers MER from 13.4\% to 12.6\% on \textsc{DevMan} and 
from 19.2\% to 18.7\% on \textsc{DevSGE}.
Both comparison keep the \SI{100}{h} ground-truth corpus fixed; the performance delta therefore isolates the benefit of
speaker and prosody diversity introduced by TTS.
Speed-perturbation merely warps temporal dynamics while preserving a single speaker identity, 
whereas our TTS pipeline injects hundreds of synthetic speaker embeddings, enriching the acoustic information to finetune Whisper. Therefore, showing the importance of speaker variety in comparison to purely speed perturbing the data only.

\subsection{Cross-Domain Comparison}
\begin{table}[t]
  \centering
    \caption{MER (\%) of Whisper-small on the ASCEND-Test set after $\sim$9 h of
           domain-specific training, comparing ground-truth finetuning with two
           CosyVoice-generated data variants.  Bold marks the best (lowest)
           MER.}
  \begin{tabular}{@{}lc@{}}
    \toprule
    \textbf{Data} & \textbf{ASCEND-Test} \\
    \midrule
    Ground-Truth          & \textbf{17.8} \\
    \midrule
    CosyVoice zero-shot gen.  & 25.2 \\
    CosyVoice (SEAME-FT) gen. & 19.1 \\
    \bottomrule
  \end{tabular}

  \label{tab:ascend_cosyvoice_vs_gt}
\end{table}

ASCEND\footnote{Corpus and license at \url{https://github.com/HLTCHKUST/ASCEND}.}
is a fully transcribed, \SI{9}{h} collection of spontaneous
Chinese–English code-switching dialogue collected in Hong Kong.
The material comprises roughly 12\,000 utterances from 38 speakers
(21 female, 17 male) with an average duration of 2.4 s per utterance,
captured at \SI{16}{kHz}.  Similar to conversational style data in SEAME,
ASCEND also reflects conversational turn-taking.

To demonstrate the portability of CosyVoice generated data, we used the SEAME-adapted CosyVoice to synthesise speech for ASCEND, a separate Chinese–English codeswitching dialogue corpus that shares the fast turn-taking and informal lexical mixing typical of everyday conversation. We input the 9 h of ASCEND text transcripts to the SEAME-tuned TTS, and regenerate the exact same utterances using TTS to see the effects of bringing SEAME-adapted CosyVoice to another domain

As summarised in Table \ref{tab:ascend_cosyvoice_vs_gt}, replacing the collected ASCEND audio with synthetic speech from an unadapted CosyVoice model (“zero-shot”) degrades Whisper-small from 17.8\% to 25.2\% MER, a 42\% relative drop that underscores how strongly ASR performance depends on the prosodic and stylistic match between training and test domains. When the very same TTS engine is first fine-tuned on SEAME, a corpus that shares ASCEND’s conversational, code-switching characteristics and then used to regenerate the ASCEND utterances, MER falls to 19.1\%. This SEAME-aligned synthetic data erases three quarters of the error penalty introduced by the zero-shot condition, cutting MER by 24\% relative and leaving only a 1.3 \% gap to the ground-truth baseline. The result demonstrates that a single round of style adaptation enables CosyVoice to produce training speech that is nearly as effective as real recordings, offering a cost-efficient path for bootstrapping ASR in new conversational code-switching domains without further data collection.

\section{Conclusions and Future Work}
We investigated multilingual text-to-speech (TTS) as a viable data augmentation technique for addressing the challenge of limited conversational code-switching data in automatic speech recognition. Our results demonstrate that fine-tuning a modern multilingual TTS model to generate synthetic speech effectively captures the diverse speakers, informal lexical choices, and spontaneous prosody typical of real-world code-switching conversations. The synthetic data produced by our method substantially increases training diversity and realism, providing a practical, cost-efficient way to enhance ASR robustness in low-resource, conversational scenarios.

However, the current augmentation pipeline is constrained by the limited textual diversity inherent in existing transcriptions. As future work, we plan to explore enhancing text variability by employing large language models (LLMs) specifically designed or adapted for multilingual and code-switching text generation. Although reliable and controlled code-switching text generators remain unavailable, their development would enable the synthesis of richer, more varied training examples, further strengthening ASR performance across diverse multilingual conversational domains.

\section*{Acknowledgment}
The computational resources for this article was performed on resources of the National Supercomputing Centre, Singapore (https://www.nscc.sg) in collaboration with Institute for Infocomm Research (I2R A*STAR Singapore)

\printbibliography

\end{document}